\documentclass[a4paper,11pt]{article}
\usepackage{pos}
\usepackage{ulem}
\usepackage{color}
\usepackage{amsfonts}
\usepackage{mathrsfs}

\title{The lower moments of nucleon structure functions in lattice QCD with physical quark masses}
\ShortTitle{The lower moments of nucleon structure functions....}

\author*[\dag, a]{Ryutaro Tsuji}
\author[b]{Yasumichi Aoki}
\author[c]{Ken-Ichi Ishikawa}
\author[d]{Yoshinobu Kuramashi}
\author[a]{Shoichi Sasaki}
\author[d]{Eigo Shintani}
\author[e.d]{Takeshi Yamazaki}
\note[\dag]{Present adress is \textit{RIKEN Center for Computational Science, 650-0047, Kobe, Japan}.}
\affiliation[]{\normalsize{\bf \sffamily \hspace{50mm}(PACS Collaboration)}}

\affiliation[a]{Department of Physics, Tohoku University,\\
  980-8578,Sendai, Japan}

\affiliation[b]{RIKEN Center for Computational Science,\\
  650-0047, Kobe, Japan}

\affiliation[c]{Core of Research for the Energetic Universe, Graduate School of Advanced Science and Engineering,\\
  739-8526, Higashi-Hiroshima, Japan}

\affiliation[d]{Center for Computational Sciences, University of Tsukuba,\\
  305-8577, Tsukuba, Japan}

\affiliation[e]{Faculty of pure and Applied Sciences, University of Tsukuba,\\
  305-8571, Tsukuba, Japan}

\emailAdd{tsuji@nucl.phys.tohoku.ac.jp}

\abstract{
    We present results for the nucleon structure functions and form factors obtained from 2+1 flavor lattice QCD with physical light quark masses ($m_{\pi}=135$ MeV) in a large spatial extent of about 10 fm. Our calculations are performed with the PACS10 gauge configurations generated by the PACS Collaboration with the six stout-smeared ${\mathscr{O}}(a)$ improved Wilson-clover quark action and Iwasaki gauge action at $\beta=1.82$ and $2.00$ corresponding to lattice spacings of $0.085$ fm and $0.064$ fm respectively. 
The lower moments of structure functions, $\langle x \rangle_{u-d}$ and 
$\langle x \rangle_{\Delta u - \Delta d}$ given by the twist-2 operators being properly renormalized, 
are evaluated in the $\overline{\rm MS}$ scheme at the renormalization scale of 2 GeV only at $\beta=1.82$, since the renormalization factors at $\beta=2.00$ have not yet determined nonperturbatively in the RI/MOM scheme. Instead, at two lattice spacings, we evaluate appropriate ratios of $g_{A}/g_{V}$ and $\langle x \rangle_{u-d}/\langle x \rangle_{\Delta u -\Delta d}$, which are not renormalized in the continuum limit. These quantities thus can be directly compared with the experimental data without the renormalization.}

\FullConference{%
 The 38th International Symposium on Lattice Field Theory, LATTICE2021
  26th-30th July, 2021
  Zoom/Gather@Massachusetts Institute of Technology
}

\begin{document}
\maketitle

\section{Introduction}
In the standard model of modern particle physics, the nucleon is known
to be a composite particle made of quarks and gluons, and their interactions are described by QCD.
Due to the nonperturbative nature of QCD at low energy scales, 
the nucleon structure that is governed by strong many body problem of the elementary constituents is one of the great challenges of lattice QCD. 
An appropriate theoretical framework to investigate the nucleon structure
is established through
the transverse-momentum-dependent parton distributions and
the generalized parton distributions, which unify the concepts of 
parton distributions obtained from the deep inelastic scattering experiments and 
nucleon elastic form factors.
The parton distributions are not only used as inputs for simulations of high
energy scattering processes in high energy physics~\cite{0}, but also play important roles to describe various properties of the nucleon such as
proton's single spin asymmetry, hadron tomography, gluons saturation and so on~\cite{1,2}.

Apart from the direct calculation of the parton distributions from lattice QCD~\cite{5-1,5-2}, it is still important to 
 determine the lower moments of the parton distributions associated with twist-2 operators.
Since the isovector momentum fraction ($\langle x \rangle_{u-d}$)
and the isovector helicity fraction ($\langle x \rangle_{\Delta u - \Delta d}$) are well evaluated from the experimental data, the calculation of 
these quantities is a benchmark to explore the nucleon structure from lattice QCD.
In addition, the precision on the isovector axial-vector coupling ($g_A$) 
is greatly improved by the current $\beta$-decay measurements with cold and ultracold neutrons~\cite{8}. 
Therefore, it offers an opportunity to pursue the high-precision calculation of $g_A$ in lattice QCD in the context of a benchmark study.

%the isovector axialvector coupling which is also measured by $\beta$-decay with cold and ultracold %neutrons quite precisely~\cite{6,7,8} is calculated. The high-precision and high-accuracy calculation %whose precision and accuracy are comparable with the experimental ones that show the %consistency of both can support the validity.

\section{Method}
\label{Sec_method}
Both the lower moments of parton distributions (Twist-2) and the axial-vector (vector) coupling (Local) can be evaluated from the nucleon matrix element of a given bilinear operator, $O_{\Gamma}=\bar{\psi}\Gamma\psi$ as defined in Table~\ref{tab:oprators}.

%
% Table
%
\begin{table*}[ht]
    \caption{The correspondence of the quantities and operators in the matrix elements\cite{10}.
\label{tab:oprators}}
\centering
\begin{tabular}{ccccc}
          \hline        
    Type of operator &  \multicolumn{2}{c}{Local}& \multicolumn{2}{c}{Twist-2}\\
    Observable & $g_V$ (vector)& $g_A$ (axial-vector) &
    $\langle x \rangle_{u-d}$ (unpolarized) & $\langle x \rangle_{\Delta u - \Delta d}$ (polarized)\\
          \hline        
          \hline        
    $\Gamma$ & $\gamma_{4}$& $\gamma_{i}\gamma_5$
            & $\gamma_4\overleftrightarrow{D}_{4}-\frac{1}{3}\sum_{k}\gamma_k\overleftrightarrow{D}_{k}$
%            & $\gamma_{\{3}\overleftrightarrow{D}_{4\}}$\\
            & $\gamma_{3}\overleftrightarrow{D}_{4}+\gamma_{4}\overleftrightarrow{D}_{3}$\\
          \hline        
\end{tabular}
\end{table*}

In general, the nucleon matrix elements are evaluated from a ratio of the nucleon three-point function with a given operator $O_{\Gamma}$ inserted at $t=t_{\rm op}$ being subject to a
 range of $t_{\rm snk} > t > t_{\rm src}$, to the nucleon two-point function with a
 source-sink separation ($t_{\rm sep}= t_{\rm snk}-t_{\rm src}$) as 
\begin{align}
    \label{eq:method_ratio}
    R(t_{\rm sep},t_{\rm op})
    \equiv
    \frac{C^{\rm 3pt}(t_{\rm op},t_{\rm sep})}{C^{\rm 2pt}(t_{\rm sep})}
    \underset{t_{\rm sep}\gg t_{\rm op}\gg 0}{\rightarrow}
    \langle 1 | O_{\Gamma} |1 \rangle
    + {\mathscr{O}}(\mathrm{e}^{-t_{\rm sep}\Delta})
    + {\mathscr{O}}(\mathrm{e}^{-(t_{\rm sep}-t_{\rm op})\Delta}),
\end{align}
where $|i\rangle$ represents the $i$-th energy eigenstate and $i=1$ stands for the ground state of the nucleon.  If the condition $t_{\rm sep}\gg t_{\rm op}\gg 0$
is satisfied, the desired matrix element $\langle 1 | O_{\Gamma} |1 \rangle$ can be read off from an asymptotic plateau, which is independent of a choice of $t_{\rm op}$.
Narrower source-sink separation causes
systematic uncertainties stemming from the excited states contamination represented by two terms of ${\mathscr{O}}(\mathrm{e}^{-t_{\rm sep}\Delta})$ and ${\mathscr{O}}(\mathrm{e}^{-(t_{\rm sep}-t_{\rm op})\Delta})$, where $\Delta\equiv E_{2}-E_{1}$ denotes 
a difference between the two energies
of the ground state ($E_1$) and the lowest excited state ($E_2$).

    In this study, the nucleon interpolating operator is constructed by
    the exponentially smeared quark operators, so as to maximize an
    overlap with the nucleon ground state as

\begin{align}
    \label{eq:interpolating_op}
    N(t,\vec{p})&
    =
    \sum_{\vec{x}\vec{x_1}\vec{x_2}\vec{x_3}}
    \mathrm{e}^{-i\vec{p}\cdot\vec{x}}\varepsilon_{abc}
    \left[
        u^{T}_{a}(t,\vec{x_1})C\gamma_5d_b(t,\vec{x_2})
    \right]
    u_c(t,\vec{x_3})
    \times 
    \Pi_{i=1}^{3}
    \phi(\vec{x_i}-\vec{x})\nonumber\\
                &\mathrm{with} \quad
    \phi(\vec{x_i}-\vec{x})
    =
    A\mathrm{exp}(-B|\vec{x_i}-\vec{x}|)
\end{align}
where there are two smearing parameters $(A, B)$. Since the condition, $t_{\rm sep}\gg t_{\rm op}\gg 0$ appearing in Eq.~(\ref{eq:method_ratio}) 
is usually not satisfied in practice, the excited states contaminations 
{could not be fully eliminated by tuning smearing parameters.

In order to eliminate the systematic uncertainties,
one should calculate the ratio (\ref{eq:method_ratio}) with several choices of $t_{\rm sep}$, and then makes sure whether the evaluated value of the nucleon matrix element does not change with a variation of  $t_{\rm sep}$ within a certain precision. This is called the ratio method that is mainly used
in this study. Alternatively, there is another method called the summation method~\cite{11}.
The summation method use a sum of the ratio $R(t_{\rm op},t_{\rm sep})$ with respect to
$t_{\rm op}$ as 
\begin{align}
    \label{eq:method_summ}
    S(t_{\rm sep})
    \equiv
    \sum^{t_{\rm sep}}_{t_{\rm op}=0}
    R(t_{\rm op},t_{\rm sep})
    \underset{t_{\rm sep}\gg0}{\rightarrow}
    \ \ 
    \mathrm{const.} + t_{\rm sep}\langle 1 | O_{\Gamma} | 1 \rangle+ {\mathscr{O}}(\mathrm{e}^{-\Delta t_{\rm sep}})
\end{align}
When more than 3 sets of $t_{\rm sep}$ are carried out in the ratio method, one can also perform 
the summation method~\cite{11}, where the matrix element $\langle 1 | O_{\Gamma} |1 \rangle$ can be read off from a slope of the linear dependence of $t_{\rm sep}$ as described in Eq.~(\ref{eq:method_summ}).
%If we find a large discrepancy between the results obtained from the ratio and summation methods, a
A difference of the central values obtained from the two methods is quoted as a systematic error on the matrix element evaluated from the ratio method in this study. 

In order to compare with the experimental values or other lattice results, the {\it bare matrix element} obtained from the above mentioned methods should be renormalized with the renormalization
constants $Z_{O_{\Gamma}}$ for each $\Gamma$ operator in 
a certain scheme. 
In this study, for the twist-2 operators, the Regularization Independent MOMentum subtraction (RI/MOM) scheme~\cite{12} is used as the intermediate scheme in order to evaluate the renormalization constants $Z_{O_{\Gamma}}$ in fully nonperturbative manner. The renormalization constants determined in the RI/MOM scheme are then converted to the $\overline{\rm MS}$ scheme at certain scale $\mu_0$ and evolved to the scale of 2 GeV using the perturbation theory. 

In general, the final result of $Z^{\overline{\rm MS}}_{O_{\Gamma}}({\rm 2 GeV})$ receives the residual dependence of the choice of the matching scale $\mu_0$.
It is true that the perturbative conversion from the RI/MOM scheme to the $\overline{\rm MS}$ scheme produces the residual $\mu_0$ dependence, but the perturbative uncertainty is not a major concern here. There are the other two sources as follows. One stems from lattice discretization errors at higher $\mu_0$, while another is originated from the nonperturbative effect that becomes relevant at lower $\mu_0$~\cite{14,15,16}.
The latter is not so serious if the spatial volume is not so large, because the lower momentums are not accessible in the smaller volume. 
However, in this study, we use $64^4$ lattice, corresponding to a linear spatial extent of 5.5 fm, that
is rather large spatial size.

In order to minimize the systematic uncertainties associated with the residual $\mu_0$-dependence, we used following two types of fitting functional forms as functions of the matching scale $\mu_0$:
\begin{align}
    f_{O_{\Gamma}}^{\mathrm{Global}}(\mu_0) = \frac{c_{-1}}{(\Lambda^{-1}\mu_0)^2} + c_0 + 
    \sum_{k > 0}^{k_{\rm max}} c_k (a\mu_0)^{2k}
\quad \mathrm{and}\quad
    f_{O_{\Gamma}}^{\mathrm{IR}-\mathrm{trunc.}}(\mu_0) =  c_0 + \sum_{k > 0}^{k_{\rm max}}  c_k (a\mu_0)^{2k}
\label{eq:ren_fit}
\end{align}
with $c_0$ being the $\mu_0$-independent value of $Z^{\overline{\rm MS}}_{O_{\Gamma}}(\mu)$
at the renormalization scale $\mu=2$ GeV.
The value of $k_{\rm max}$ is determined by a $\chi^2$ test for goodness fit, so that $k_{\rm max}=3$ is chosen for the former, while $k_{\rm max}=1$ is chosen for the later.
In the former form, $\Lambda$ is responsible 
for a scale associated with the nonperturbative effect. The former is applied for fitting all data, 
while the latter is used for fitting the data in a restricted range of $\mu_0 \ge \mu$.
The discrepancy between the values of $c_0$ extracted from 
%two types of 
these fittings is quoted as a systematic error on the renormalization constant $Z^{\overline{\rm MS}}_{O_{\Gamma}}({\rm 2\;GeV})$.  }

\section{Simulation details}
We mainly use the PACS10 configurations generated by the PACS Collaboration with the six stout-smeared ${\mathscr{O}}(a)$ improved Wilson-clover quark action and Iwasaki gauge action at $\beta=1.82$ and $2.00$ corresponding to the lattice spacings of $0.085$ fm (coarser) and $0.064$ fm (finer) respectively~\cite{17,18,19,20,21}. 
When we compute nucleon two-point and three-point functions, the all-mode-averaging (AMA) technique\cite{24} is employed in order to reduce the statistical errors significantly without increasing
%TSUJIthe 
computational costs. Two of three lattice ensembles are created with 
%TSUJIthe
same lattice cutoff, but on 
%TSUJIthe
different lattice sizes.
The smaller volume ensembles are used for the finite volume study on the axial-vector coupling $g_A$ and nucleon elastic form factors, and also used for computing the renormalization constants which are known to be less sensitive to the finite volume effect.
% for final values using the larger volume ensembles of $128^4$ lattice. 

\begin{table*}[ht]
\caption{Summary of simulation parameters 
used in this study.
%in 2+1 flavor PACS ensembles. 
%with two different spatial extent and cut-off.
\label{tab:simulation_details}}
\begin{tabular}{cccccccc}
          \hline        
&  $\beta$ &$L^3\times T$ & $a^{-1}$ [GeV] & $La$ [fm]& $\kappa_{ud}$ & $\kappa_{s}$& $M_\pi$ [GeV] \\
          \hline        
          \hline        
    $128^4$ lattice& 1.82& $128^3\times 128$& 2.3162(44)& 10.9&0.126117 & 0.124902 &  0.135\\
          \hline
    $64^4$ lattice& 1.82& $64^3\times 64$   & 2.3162(44)& 5.5 &0.126117  &0.124902  & 0.138 \\
          \hline
    $160^4$ lattice& 2.00& $160^3\times 160$ & 3.0892(25)& 10.3 &0.12584  &0.124925  & 0.135 \\
          \hline
\end{tabular}
\end{table*}

\begin{table*}[ht]
    \caption{Details of the measurements: time separation 
    ($t_{\rm sep}$), the smearing parameters $(A,B)$, the number of high-precision calculation 
    ($N_{\mathrm{org}}$), the number of configuration ($N_{\mathrm{conf}}$), the 
    measurements per configuration ($N_{G}$) and the total number of the measurement 
    ($N_{\mathrm{meas}}=N_{\mathrm{conf}} \times N_{G}$), respectively. 
\label{tab:measurements}}
\centering
\begin{tabular}{ccccccccc}
          \hline
&  $t_{\rm sep}/a$ &Smearing parameters& $N_{\mathrm{org}}$&  $N_{G}$ & $N_{\mathrm{conf}}$ & $N_{\mathrm{meas}}$\\
\hline
          \hline
    $128^4$ lattice& 10&$(A,B)=(1.2,0.16)$ & 1& 128& 20& 2,560\\
                   & 12&             & 1& 256& 20& 5,120\\
                   & 14&             & 2& 320& 20& 6,400\\
                   & 16&             & 4& 512& 20& 10,240\\
          \hline
    $64^4$ lattice& 12&$(A,B)=(1.2,0.14)$ & 4& 256& 100& 25,600\\
                  & 14&             & 4& 1,024& 100& 102,400\\
                  & 16&             & 4& 2,048& 100& 204,800\\
          \hline
    $160^4$ lattice& 16&$(A,B)=(1.2,0.11)$ & 4& 64& 15& 9,210\\
                   & 19&                    & 2& 64& 15& 15,360\\
          \hline
\end{tabular}
\end{table*}

\section{Numerical results}
In this study, we first present the preliminary results for the renormalized values of the isovector momentum fraction ($\langle x \rangle_{u-d}$) and the isovector helicity fraction ($\langle x \rangle_{\Delta u - \Delta d}$), which are calculated only at a single lattice spacing with the $128^4$ lattice ensemble. We will later discuss the discretization uncertainties on two specific ratios of $g_{A}/g_{V}$ and $\langle x \rangle_{u-d}/\langle x \rangle_{\Delta u -\Delta d}$, which are not renormalized in the continuum limit. These quantities, which are evaluated with both $128^4$ and $160^4$ lattice ensembles,
can be directory compared with the experimental data without renormalization. 

\subsection{Renormalized values of momentum and helicity fractions from $128^4$ lattice ensemble}
In order to evaluate 
$\langle x \rangle_{u-d}$ and $\langle x \rangle_{\Delta u - \Delta d}$,
we calculate both the bare nucleon matrix elements of the relevant twist-2 operators and their renormalization constants at lattice spacing of $a=0.085$ fm. 
As for the calculations of the bare matrix elements, 
four data sets using $t_{\rm sep}/a=\{10,12,14,16\}$ are calculated and two methods are employed in the extraction of the bare matrix elements. According to the analysis based on the ratio method, we first observed that $t_{\rm sep}/a=\{14,16\}$ for the unpolarized case and $t_{\rm sep}/a=\{12,14,16\}$ for the polarized case are large enough 
for reducing the excited states contamination in three-point functions with the twist-2 operators respectively. Next, all data sets with four values of $t_{\rm sep}/a$
are used in the summation method in order to estimate 
the systematic error associated with the excited states contamination in each channel.

The renormalization constants of the local vector and axial-vector currents are obtained with the Schr\"odinger functional scheme at vanishing quark masses as $Z_{V}=0.95153(76)(1487)$ and $Z_{A}=0.9650(68)(95)$~\cite{Ishikawa:2015fzw}, while the renormalization constants of the twist-2 operators are evaluated in the RI/MOM scheme as described in Sec.~\ref{Sec_method}.
%Furthermore, their renormalization constants are calculated using the renormalization constants of local vector and axial couplings, $Z_{V}=0.95153(76)(1487)$ and $Z_{A}=0.9650(68)(95)$, which are obtained with the Schr\"odinger functional scheme, and are extracted by the two fitting functions of Eq.~\ref{eq:ren_fit}. 

 \begin{table}[ht]
     \caption{List of error sources in
         the renormalized nucleon 
     matrix elements with the twist-2 operators.  }
     %Error sources and their precision contributions our preliminary results of the twist-2 operators. %In addition, ones of the total error including both statistical and systematic errors of renormalized values also are summarized at the bottom of this table.}
     \label{tab:err_src}
     \centering
     \begin{tabular}{ccccccc}
       \hline
                       & \multicolumn{6}{c}{List of errors (\%)} \\
       \cline{2-7}
             & \multicolumn{3}{c}{Momentum fraction $\langle x \rangle_{u-d}$}
             & \multicolumn{3}{c}{Helicity fraction $\langle x \rangle_{\Delta u-\Delta d}$}\\
         Error source & Total &  Bare value & $Z$-factor
                         & Total &  Bare value & $Z$-factor\\
       \hline
       \hline
         Statistical  &  5.30  & 4.60 & 2.52
                          &  4.39  & 3.49 & 2.91\\
         Systematic &  31.1  & 19.5 & 24.5
                          &  24.0  & 3.49 & 22.5 \\
       \hline
     \end{tabular}
   \end{table}

\begin{figure*}
 \centering
 \includegraphics[width=0.78\textwidth,bb=50 45 752 552, clip]{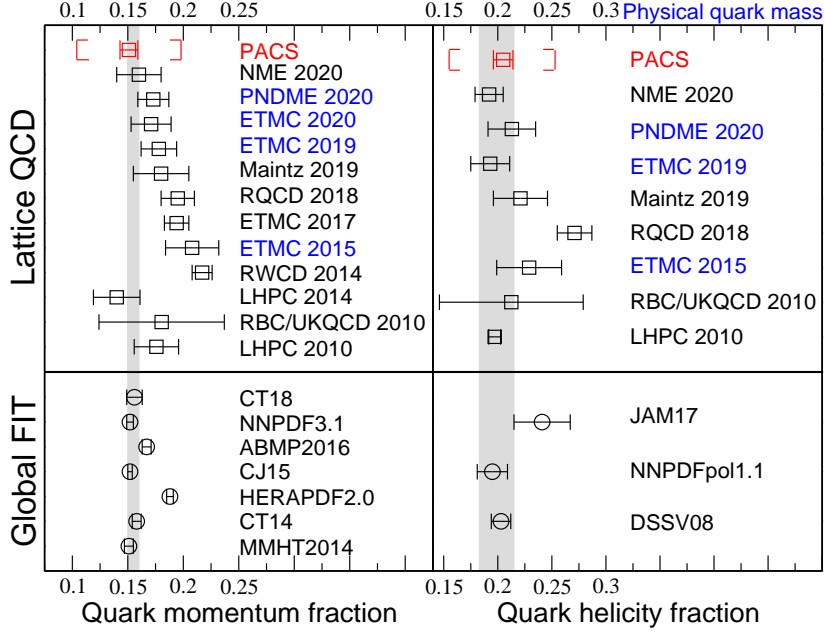}
 \caption{Comparison of recent lattice QCD calculations (upper panels) and global QCD analyses of experimental data (lower panels). The red symbols represent our preliminary results. 
The error bars denote their statistical errors only, while the outer brackets with our results represent the total errors including the systematic one. The gray bands correspond to 
the values of an average given by global fit determinations~\cite{5-1,5-2}.
}
 \label{fig:twist2_comp}
\end{figure*}

As shown in Fig~\ref{fig:twist2_comp}, we compare our preliminary results of
the renormalized values of $\langle x \rangle_{u-d}$ (left panels) and $\langle x \rangle_{\Delta u-\Delta d}$ (right panels) with the values obtained from recent lattice QCD calculations (upper panels)
and global QCD analyses of experimental data (lower panels). The central values of our results are well within gray bands corresponding to the experimental
values 
given by global fit determinations~\cite{5-1,5-2}: $\langle x \rangle_{u-d} = 0.155(5)$ and $\langle x \rangle_{\Delta u -\Delta d} = 0.199(16)$. It is worth mentioning
that our statistical precisions are also comparable to the global QCD analyses. This indicates that lattice QCD
is partially qualified to directly investigate the parton distributions
as an alternative to actual experiments. However, it should be kept in mind that our preliminary results
of $\langle x \rangle_{u-d}$ and $\langle x \rangle_{\Delta u-\Delta d}$ still have large systematic errors as shown in Tab.~\ref{tab:err_src}.
As for the bare matrix elements, discrepancies between the results obtained from the two
methods are quoted as 
a systematic error stemming from the excited states contaminations.
It is found that for the unpolarized case the evaluated systematic error with the present sets of 
$t_{\rm sep}/a=\{10,12,14,16\}$ is larger than the statistical one, while the statistical and systematic uncertainties are also comparable in the polarized case. A further calculation with the larger choice of $t_{\rm sep}/a(>16)$ is required to reduce this particular uncertainty. 
The largest error on $\langle x \rangle_{u-d}$ ($\langle x \rangle_{\Delta u-\Delta d}$)
is the systematic uncertainty in determination of the renormalization constants.
This systematic error was evaluated by a difference between the results obtained 
by two fitting procedures with Eqs.~(\ref{eq:ren_fit})
which have different ways
to treat the residual $\mu_0$-dependence in
smaller $\mu_0$ region.
As shown in Tab.~\ref{tab:err_src}, the total systematic errors, which are about 6 times larger than 
the total statistical errors, are dominated by the systematic errors on the renormalization constants.
To reduce the total systematic error, the RI/SMOM scheme
that was used for determination of the renormalization constant 
for the local operators~\cite{Tsukamoto:2019dmi}, could be useful even for the twist-2 operators.

\subsection{Ratios of $g_{A}/g_{V}$ and $\langle x \rangle_{u-d}/\langle x \rangle_{\Delta u -\Delta d}$ from $128^4$ and $160^4$ lattice ensembles}

\begin{figure*}
 \includegraphics[width=0.49\textwidth,bb=24 30 752 532, clip]{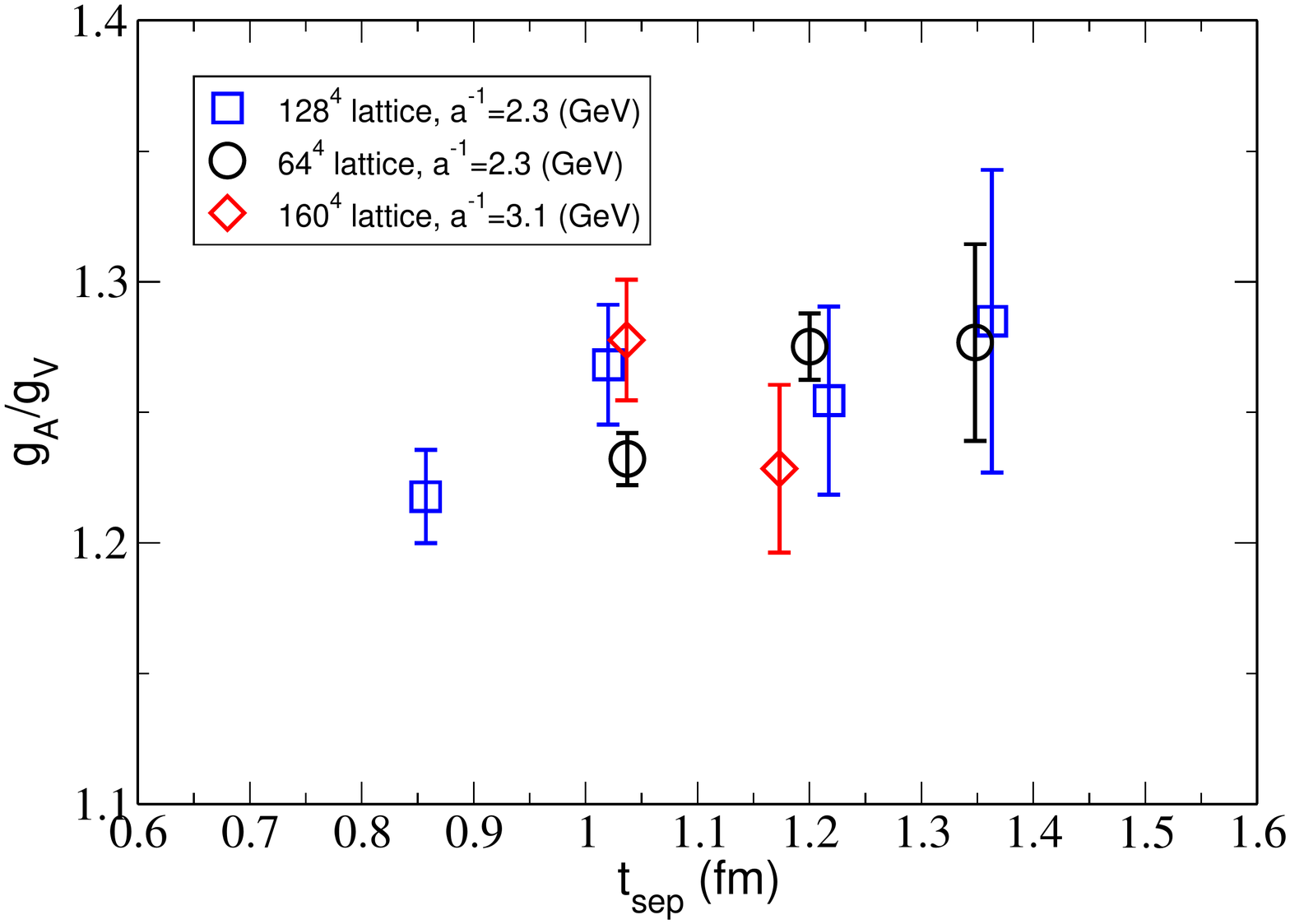}
 \includegraphics[width=0.49\textwidth,bb=24 30 752 532, clip]{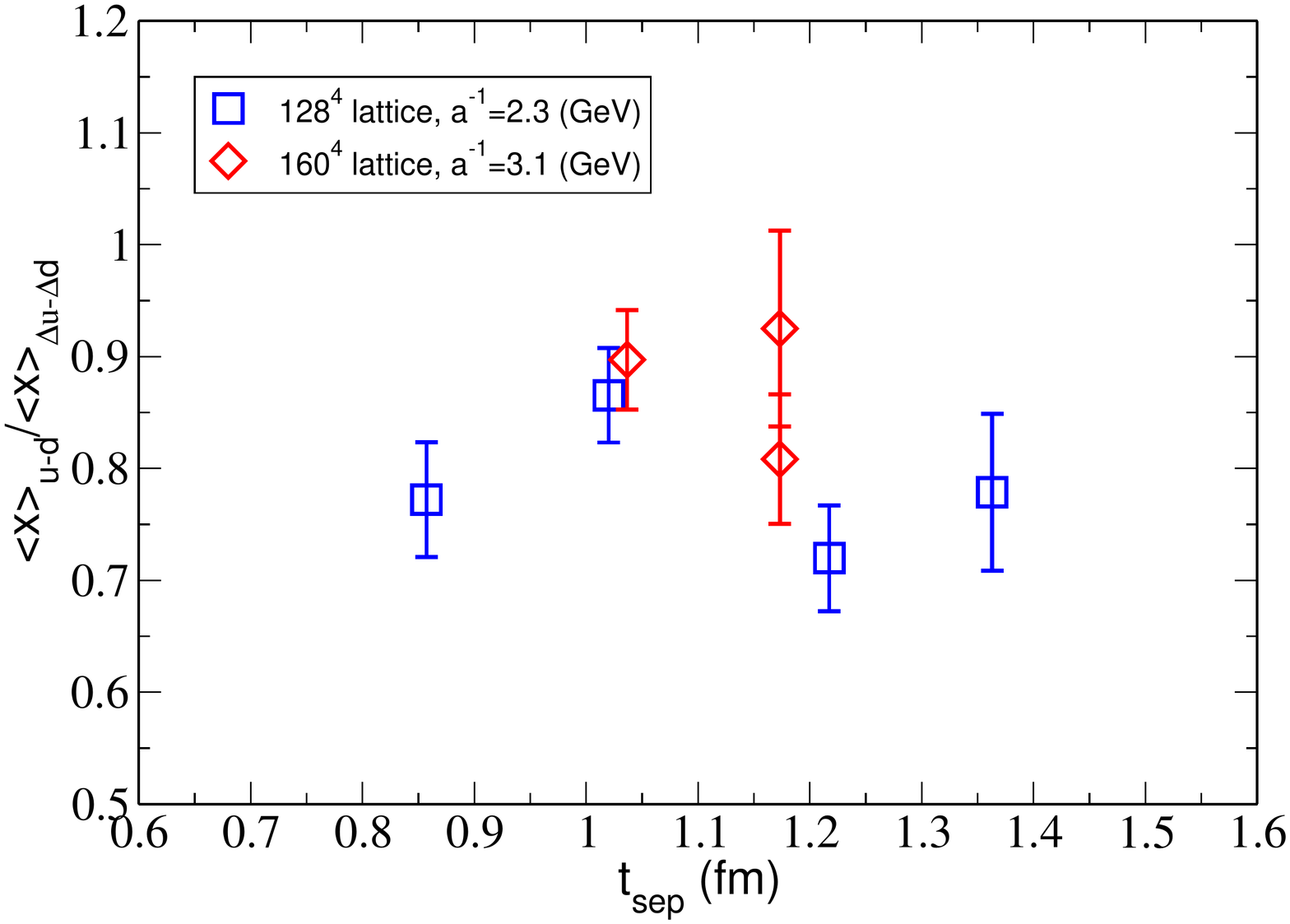}
 \caption{Ratios of bare values of $g_{A}$ and $g_{V}$ (left), and $\langle x \rangle_{u-d}$ and $\langle x \rangle_{\Delta u -\Delta d}$ (right) as a function of $t_{\rm sep}$. }
 \label{fig:continuum_esc}
\end{figure*}

\begin{figure*}
 \centering
 \includegraphics[width=0.78\textwidth,bb=20 45 752 525, clip]{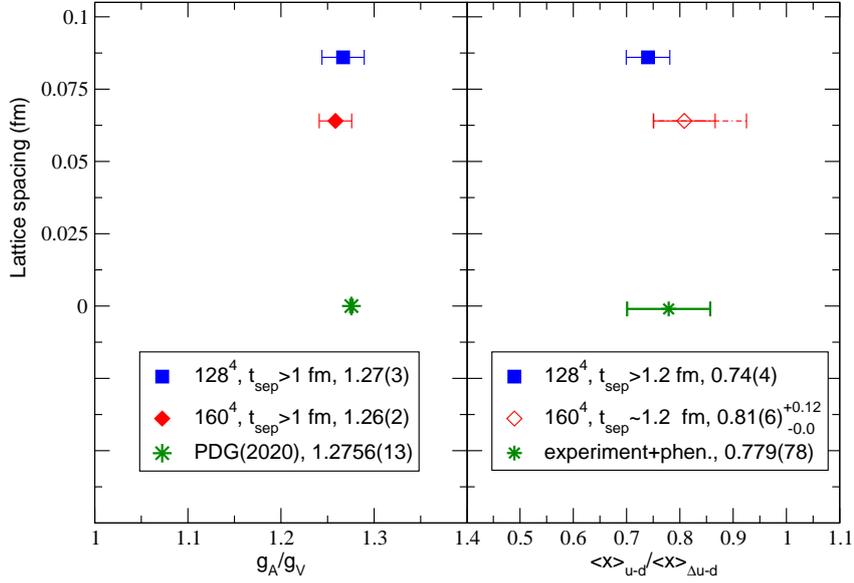}
 \caption{Preliminary results of $g_{A}/g_{V}$ and $\langle x \rangle_{u-d}/\langle x \rangle_{\Delta u -\Delta d}$ obtained from $128^4$ and $160^4$ lattice ensembles. 
The value of $\langle x \rangle_{u-d}/\langle x \rangle_{\Delta u -\Delta d}$ obtained from the
$160^4$ lattice ensemble (open diamond symbol) receives a large uncertainty regarding the fit range dependence
%, whose size is represented by the length of the dashed error bar.
that is represented by dashed error bar.
}
 \label{fig:continuum_comp}
\end{figure*}

We next present our preliminary results of $g_{A}/g_{V}$ and $\langle x \rangle_{u-d}/\langle x \rangle_{\Delta u -\Delta d}$ obtained from both $128^4$ and $160^4$ lattice ensembles,
since the renormalization constants are not yet evaluated at finer lattice spacing.
In Fig~\ref{fig:continuum_esc}, we first shows the $t_{\rm sep}$ dependence of the ratios of $g_{A}/g_{V}$ (left panel) and $\langle x \rangle_{u-d}/\langle x \rangle_{\Delta u -\Delta d}$ (right panel), both of which
are evaluated with the ratio method. 
All three ensembles that include the $64^4$ lattice ensemble are used to 
calculate the values of $g_{A}/g_{V}$, while only the $128^4$ and $160^4$ lattice ensembles
are used to calculate the value of $\langle x \rangle_{u-d}/\langle x \rangle_{\Delta u -\Delta d}$.

At first glance, the conditions of $t_{\rm sep}\ge1$ fm for the case of $g_{A}/g_{V}$
and $t_{\rm sep}\ge1.2$ fm for the case of $\langle x \rangle_{u-d}/\langle x \rangle_{\Delta u -\Delta d}$ 
can be read off from the $128^4$ lattice results, in order to keep the systematic uncertainties stemming from the excited states contamination smaller than the statistical ones.
As for the results of $g_{A}/g_{V}$, the preliminary results 
from the $160^4$ lattice ensemble show a consistent observation,
though the higher accuracy results from the $64^4$ lattice ensemble
reveal that the data of $t_{\rm sep}\approx 1$ fm is slightly deviated from the results for 
$t_{\rm sep}\ge1.2$ fm.

On the other hand, the preliminary results of $\langle x \rangle_{u-d}/\langle x \rangle_{\Delta u -\Delta d}$ from the $160^4$ lattice ensemble have not shown any obvious trend of $t_{\rm sep}$ dependence, since a large uncertainty still remains for the data given with $t_{\rm sep}\approx 1.2$ fm. 
As shown in the right panel of Fig~\ref{fig:continuum_esc}, two data points, which are obtained from 
two fitting ranges used in the ratio method, are plotted at $t_{\rm sep}\approx 1.2$ fm.
The upper data point agrees the data obtained with $t_{\rm sep}\approx 1.0$ fm, while the lower 
data point follows the trend observed in the $128^4$ lattice results.
It is difficult to draw any firm conclusion whether the systematic uncertainties stemming from
the excited states contamination are under control for the $160^4$ lattice results 
of $\langle x \rangle_{u-d}/\langle x \rangle_{\Delta u -\Delta d}$ within the current statistics. 
Therefore, in the later discussion, we use the lower data point of $t_{\rm sep}\approx 1.2$ fm for the $160^4$ lattice result, while a systematic error estimated by a difference between two data points
is also quoted.

We next examine the lattice discretization uncertainties on $g_A/g_V$ and $\langle x \rangle_{u-d}/\langle x \rangle_{\Delta u -\Delta d}$ using the results obtained at two lattice spacings as shown in Fig~\ref{fig:continuum_comp}.
According to what we discussed using Fig~\ref{fig:continuum_esc}, in the case of $g_A/g_V$ 
there is no appreciable $t_{\rm sep}$ dependence in the large volume results obtained
from 
%the $128^4$ and $160^4$ 
both
lattice ensembles if $t_{\rm sep}\approx 1.0$ fm is satisfied. 
Therefore, we simply use the combined results with $t_{\rm sep}/a=\{12, 14, 16\}$ for the $128^4$
lattice and $t_{\rm sep}/a=\{16, 19\}$ for the $160^4$ lattice in the left panel of Fig~\ref{fig:continuum_esc}. Both results at two lattice spacings
reproduce the experimental value~\cite{8} within their statistical errors. 
This indicates that the lattice discretization uncertainties on the renormalized value of 
$g_A$ are smaller than their statistical uncertainty of less than 2\%.

As for the quantity of $\langle x \rangle_{u-d}/\langle x \rangle_{\Delta u -\Delta d}$, 
although the $128^4$ lattice result obtained at the coarser lattice spacing
shows better agreement with the experimental values~\cite{5-1,5-2}, the large systematic error
remains for the preliminary result obtained from the $160^4$ lattice ensemble.
Needless to say, further reduction of both statistical and systematic errors is needed for the 
$160^4$ lattice results.

\section{Summary}

We have calculated the renormalized values of $\langle x \rangle_{u-d}$ and $\langle x \rangle_{\Delta u - \Delta d}$ at a single lattice spacing of 0.085 fm, and also the two specific ratios of $g_{A}/g_{V}$ and $\langle x \rangle_{u-d}/\langle x \rangle_{\Delta u -\Delta d}$ at two lattice spacings of 0.085 fm 
and 0.064 fm , using the PACS10 gauge configurations. This work is a benchmark to explore the nucleon structure from lattice QCD calculations. 
First, we have succeeded in reproducing global QCD analyses of experimental 
data for both quantities of $\langle x \rangle_{u-d}$ and $\langle x \rangle_{\Delta u - \Delta d}$ with high statistical precision
using the $128^4$ lattice ensemble. However, it is worth remarking that the large systematic 
uncertainties remain in determination of the renormalization constants.
In 
future project, instead of the RI/MOM scheme, the RI/SMOM scheme
will be able to be used even for the twist-2 operators. Since the renormalization constants are not yet evaluated 
at the finer lattice spacing, we simply evaluate two ratios of $g_{A}/g_{V}$ and $\langle x \rangle_{u-d}/\langle x \rangle_{\Delta u -\Delta d}$, which are not renormalized in the continuum limit.
Therefore, we can compare our results of the ratios with the experimental results without the renormalization constants. As for the case of $g_{A}/g_{V}$, both results at two lattice spacings 
well reproduce the experimental value within their statistical errors. This indicates that the lattice 
discretization uncertainties on the renormalized value of $g_{A}$ are smaller than their statistical uncertainty of less than 2\%. As in the case of $g_{A}/g_{V}$, the results of $\langle x \rangle_{u-d}/\langle x \rangle_{\Delta u -\Delta d}$ obtained from the $128^4$ and $160^4$ lattice ensembles
reproduce the experimental data, though the large systematic error remains for the preliminary 
result of obtained from the $160^4$ lattice ensemble.

\section*{Acknowledgement}
Numerical calculations in this work were performed on Oakforest-PACS in Joint Center for Advanced High Performance Computing (JCAHPC) and Cygnus in Center for Computational Sciences at University of Tsukuba under Multidiscilinary Cooperative Research Program of Center for Computational Sciences, University of Tsukuba, and Wisteria/BDEC-01 in the Information Technology Center, The University of Tokyo. This research also used computational resources through the HPCI System Research Projects (Project ID: hp170022, hp180051, hp180072, hp180126, hp190025, hp190081, hp200062, hp200188, hp210088) provided by Information Technology Center of the University of Tokyo and RIKEN Center for Computational Science (R-CCS). The  calculation employed OpenQCD system\footnote{http://luscher.web.cern.ch/luscher/openQCD/}. This work was supported in part by Grants-in-Aid for Scientific Research from the Ministry of Education, Culture, Sports, Science and Technology(Nos. 18K03605, 19H01892).

\end{document}